\documentclass[12pt]{article}
\usepackage{openwork}
\usepackage{amssymb}
\usepackage{amsthm}
\usepackage{hyperref}
\title{Confidence as Forecast: A Decision-Theoretic Interpretation of Confidence Intervals}
\author[1]{Scott Lee}
\date{}
\affil[1]{National Center for Emerging and Zoonotic Infectious Diseases, Centers for Disease Control and Prevention}
\newtheorem{theorem}{Theorem}[section]

\begin{document}

\maketitle

\begin{abstract}
What, if anything, should a frequentist say about a single realized confidence interval (CI) and its chance of having covered the parameter? Jerzy Neyman's original answer was to refuse any nondegenerate probability for coverage ex post and, instead, to ``state that the interval covers.'' In this paper I argue that the usual frequentist machinery already supports a different reading. I treat the coverage event as a Bernoulli random variable, with the nominal level $1-\alpha$ as its design-based success probability, and view ``confidence'' as a probability forecast for that Bernoulli outcome. Using strictly proper scoring rules, I show that $1-\alpha$ is the unique optimal constant forecast for coverage, both before and after observing the data, and that it remains optimal post-trial in common unbounded, translation-invariant models with pivot-based CIs. When the design yields a $\theta$-free statistic---such as the relative width of the interval in a finite-window uniform model---the conditional coverage given that statistic provides a nonconstant, design-based refinement of $1-\alpha$ that strictly improves predictive performance. Two thought experiments, a Monty Hall–style shell game and the ``lost submarine'' example of Morey et al.\ (2016), illustrate how this perspective resolves familiar interpretational puzzles about CIs without appealing to priors or single-case subjective degrees of belief. I conclude with simple ``what to do when you see an interval'' guidance for applied work and some implications for teaching confidence intervals as tools for forecasting long-run coverage.


\textbf{Disclaimer}: The findings and conclusions in this report are those of the author and do not necessarily represent the official position of the Centers for Disease Control and Prevention.
\end{abstract}

\section{1 Introduction}

\subsection{1.1 Background}

If you were being scored on predicting whether a single confidence interval (CI) has covered its target parameter, what number would you report? Jerzy Neyman, the inventor of the CI procedure, suggested proceeding in two steps \cite{neyman1937}: first, refusing to assign any probability to coverage itself, since, on the assumption that \(\theta\) is a fixed constant and not a random variable, coverage becomes fully determined once an interval has been constructed; and second, stating that the constructed interval covers the parameter. Although the latter suggestion is not typically thought of as a forecast \textit{per se} for individual coverage events, it is essentially just that---by stating that the intervals always cover, we are issuing a constant forecast for \(P(Cover)\) at \(1\), even if we choose not to interpret the forecasting subjectively, e.g., as our personal degree of belief in whether coverage occurred. The forecast has a natural frequentist interpretation in that, under repeated sampling, it will be wrong no more than \(\alpha\%\) of the time, which Neyman sensibly appeals to this as a strength of CI theory as a means for controlling the practicing statistician's error rate in making such statements.

Two other critical facts are, of course, also true. First, confidence procedures (CPs) can be constructed from data that carry no information at all about \(\theta\), and ex ante coverage probability can appear to change substantially ex post after an interval has been constructed \cite{dasgupta2010ancillary, welch1939confidence}, leading to the standard argument that the only way to think about CIs coherently is in terms of their long-run coverage properties. Second, and perhaps more importantly, for any given interval \(I\), its coverage probability is degenerate in \(\{0,1\}\), conditioned on the realized values of its endpoints, as Neyman's point above makes clear. This is the line we most often hear about how to interpret a given CI ex post (after construction)--that the interval either does or does not cover the parameter, and that we can make no probabilistic statement about its coverage because of it. 

The "it either covers, or it doesn't" statement has been the source of confusion, consternation, and frustration for beginning statistics students for a long time, perhaps since it was first made, and has led to a number of not-quite-satisfying claims about how to interpret CIs in the applied literature. Introductory papers and applied practitioners occasionally claim that constructed intervals retain their nominal coverage probability \cite{masson2003using, hoekstra2014robust}, critiques in response claim that they retain none \cite{morey2016fallacy, greenland2016statistical, hawkins2021use, sedgwick2014understanding}, and still others suggest what intervals are trying to estimate are the interval endpoints themselves, rather than $\theta$ \cite{naimi2020can}. More often than not, the ensuing debates--which can be rather spirited \cite{morey2016continued, seidenfeld1981after, mayo1981defense}--end at philosophical impasses, with frequentists and Bayesians each appealing to foundational commitments that the other side does not share. . 

In the sections below, I hope to offer a principled frequentist reading of the concept of "confidence" and its associated intervals that aims to resolve some of this confusion (see also \cite{lee2026either} for a broader conceptual argument about ex-post probability statements) . More precisely, I treat coverage probability as having three layers, rather than a single one: the first being the event-level degenerate conditional in $\{0,1\}$ determining whether an interval covers; the second being the design-level coverage guarantee of \(1-\alpha\) that averages those conditionals over the randomness in \(X\); and the third, or the notion of "confidence", being a predictive probability, or model-based forecast, of empirical coverage with whatever information the statistician may have at hand. Under this view, we have clear bounds on what we can say about a particular interval ex post (e.g., we might predict that it covers \(\theta\) with probability \(1-\alpha\)), and, in some cases, we also have mathematical justification for updating our forecast in light of relevant features of the realized interval, for example, when we see the "trivial" interval $[-\infty,\infty]$, where we would very sensibly switch our prediction to $1$, since coverage is certain. Separating our probabilistic forecasts from the design-level coverage guarantee helps dissolve some complaints of CIs as being uninterpretable (with respect to the coverage event and not the actual value of \(\theta\)), and, as I show below, it generally respects the frequentist machinery Neyman used to build his theory (even if he might not have intended it to be used quite that way).

\subsection{1.2 Paper overview}

The remainder of the paper is organized as follows. Section 2 presents a thought experiment showing that design-level coverage probability can in fact be used to guide decision-making based on constructed intervals ex post. Section 3 introduces the relevant standard formalisms for frequentist statistical inference, and it formalizes the notion of "confidence" as a predictive probability, or model-based estimate, of interval coverage that applies both ex ante and ex post. Section 4 shows how this predictive notion might apply to CI-based inference by revisiting a commonly-cited thought experiment designed to disprove the notion that confidence has any meaning ex post, and Section 5 concludes with a recap of the main results from earlier sections; a discussion of pedagogical strategies for teaching both CI theory and its applications; and some suggested potential directions for future research. 

\section*{2 A thought experiment: "Monty's Hell"}

\subsection{2. 1 The setup}

An uncommonly truthful street performer is running an interesting variation of a shell game he calls “Monty’s Hell”. You’re pretty good with probability, and you also like to gamble, so you ask him how to play. Here are his instructions:

\textit{“Take a look at these 3 plastic cups. Under each one, I have placed a handwritten note indicating a range of dollar amounts—for example, \$10 to \$20. On a separate sheet of paper that you can’t see, I've written down a single dollar amount that falls under one, and only one, of the ranges hidden by the cups. I will shuffle the cups for a while and then let you pick one to overturn. I’ll then remove one of the cups you didn’t pick that I know does not contain the winning range; if I have to choose between two such cups, you’ll flip a fair coin (provided by me) to decide which one I remove. I’ll then give you the opportunity to switch your original choice. After you make your final choice, if the range on the paper beneath your cup contains the hidden amount on my other sheet, I’ll pay you the hidden amount; if not, you’ll pay me half of the hidden amount. The game costs \$5 to play, and the hidden amount is at least \$10—are you in?”}

Five dollars seems like a small amount to pay for the chance to win some extra cash, and you like that you’ll only pay half of the hidden amount if you lose, so you agree to play a single game.

\textit{The performer shuffles the cups, and you choose one that you think is good}. \textit{Without turning the cup over, what is your expected payout?}

Since you know nothing about either the ranges under the cups or the hidden value that one of them contains, you effectively will be choosing at random. With an initial buy-in of \(\$5\), your expected payout \(M \) will be
\begin{equation} 
\begin{split}
E[M] = P(S) \, v - (1 - P(S)) \cdot \frac{1}{2} v - 5 \
= \frac{1}{3} v - \frac{2}{3} \cdot \frac{1}{2} v - 5 \\
= \frac{1}{3} v - \frac{1}{3} v - 5 \\
= -5
\end{split}
\end{equation}
where \(v\) is the dollar amount of the hidden prize. The expected payout would be the same for any value of the hidden amount, but you appreciate that because it’s at least \(\$10\), you’ll have a \(1/3^{rd}\) chance of doing no worse than doubling your original buy-in of \$5. 

\textit{True to his promise, the performer removes one of the two cups you didn’t choose, saying that it did not cover the piece of paper with the winning range. Assuming he did not lie, what is your expected payout now? }

None of the cups has been overturned, and so you still have no information about the range in which the winning value might lie. You remember the Monty Hall paradox, though, and note that even though a losing cup has been removed from the table, and now either your cup or the remaining cup must contain the winning range, there’s still only a \(1/3\) probability that your cup will win; the remaining \(2/3\) probability of winning now belongs to the single remaining cup. If you stay with your original choice, the expected payout is the same as it was at the beginning of the game: a \(\$5\). loss If you switch to the remaining cup, however, the expected payout \(M_{switch}\) becomes
\begin{equation} 
\begin{split}
E[M_{\text{switch}}] = (1 - P(W))v - P(W) \cdot \frac{1}{2}v - 5 \\
= \left(\frac{2}{3}\right)v - \left(\frac{1}{3}\right) \cdot \frac{1}{2}v - 5 \\
= \left(\frac{2}{3} - \frac{1}{6}\right)v - 5 \\
= 0.5v - 5.
\end{split}
\end{equation}
You know the minimum value of \(v\) is \(\$10\), and so your expected payout if you switch is at least \(\$0\)—in other words, in the long run, if you switch, you will at least break even. Because of this, you switch.

\textit{Finally, the street performer lets you turn over your new cup to see its hidden range, and he turns over your original cup so you can see its range, as well. You see that the range under your cup is \$30 to \$50, and the range under your original cup was \$10 to \$29. Has your expected payout changed?}

Somewhat counterintuitively, you realize the expected payout is actually still the same, even though you’ve now seen both of the remaining hidden ranges. By the rules of the game, you now know that the hidden value must lie somewhere between \(\$10\) and \(\$50\)--the performer removed one of the losing cups, effectively pinning the prize amount to the union of the intervals in the ones remaining--but you still do not know whether your cup is the winning one, and so the expected payout, which depends only on your probability of choosing correctly and the associated rewards and penalties, has not changed—given that you switched, it is still no less than \(\$0\).

\textit{The street performer reveals the hidden amount to be \$50},\textit{ which was included in the range under your chosen cup. Congratulations—you’ve multiplied your original investment of \$5 by 9 and now have \$45. A good return by any account!}

\subsection{2.1 The parallels with CIs }

This thought experiment was designed as a direct analog to CIs, and there are two main parallels to note. First is that, as with CIs capturing \(\theta\), each cup winning is decided solely by whether its interval captures the hidden dollar amount. Second, we have that the winning value, like \(\theta\), is a fixed-but-unknown constant, with no probability distribution being placed over its value. These two facts fix whether a cup wins or loses, but, as with CIs, we have no way of knowing which. The uncomfortable dissonance here is that from both Bayesian and frequentist simulation-based analyses of the original Monty Hall problem under its standard setup, we know that switching is, indeed, the optimal strategy. Does Neyman's interpretation also lead to success? The answer in this case is "no": if we follow either of the options he gives us for interpreting CIs after choosing our first cup, we are guaranteed to lose money. The degenerate conditional in \(\{0, 1\}\) would freeze us in our tracks, since we now have no basis for weighing whether to switch and thus decide to stay; and stating "this interval covers" would also do the same, since we now have a clear reason \textit{not} to switch, as we are treating our first cup as the one that wins. In both cases, the long run would keep the chances of winning at $1/3$, and we would lose an average of $\$5$ per game, rather than taking advantage of the $2/3$ chances offered by the switch. In other words, the two Neyman-style moves—refusing to assign a nondegenerate probability to coverage, or simply declaring that our original cup “covers”—are not only philosophically awkward; rather, they are strictly outperformed by the strategy that treats the design-level success probability as a forecast. 

\section{3 Neyman through the looking-glass}

Neyman's interpretation looks at realized CIs from one point of view, but, as we saw informally above, the underlying model offers us another one. At the level of the model, these correspond to two ways of viewing the coverage variable: as a $\{0,1\}$--valued random variable attached to a specific interval, and as its nondegenerate mean $1-\alpha$ under the sampling distribution. Both are features of the same probability model. They differ only in the $\sigma$–fields with respect to which they are evaluated, not in whether they are ``pre--data'' or ``post--data'' in any temporal sense (this particular distinction is more about the choice of reference class than about what the model does or does not allow \cite{hajek2007reference}). Below, I begin by revisiting how these two layers of probability are connected by the model governing coverage, and I then formalize the third layer I suggested at the start of the paper: ``confidence'' as a predictive probability for empirical coverage, conditional on whatever information we happen to have on hand, with proper scoring rules providing guidance on what numbers to report, and when.

\subsection{3.1 What the model gives us}
\label{subsec:two-povs}
We start with a standard frequentist model
\[
(\Omega,\mathcal{F},\{P_\theta : \theta \in \Theta\}),
\]
where $\theta \in \Theta$ is a fixed but unknown parameter. The data $X$ is a random element
\[
X : (\Omega,\mathcal{F},P_\theta) \to (\mathcal{X},\mathcal{A})
\]
with distribution $P_\theta^X$. A (two–sided) $1-\alpha$ confidence interval (CI) procedure for $\theta$ is a measurable map
\[
I : \mathcal{X} \to \mathcal{I}, 
\qquad
x \mapsto I(x) = [L(x),U(x)],
\]
such that, for every fixed $\theta \in \Theta$,
\begin{equation}
  P_\theta\big(\theta \in I(X)\big) 
  \;=\;
  P_\theta\big(L(X) \le \theta \le U(X)\big)
  \;=\;
  1-\alpha.
  \label{eq:coverage-def}
\end{equation}
Next, we define the corresponding \emph{coverage indicator}
\[
Z(X) 
\;:=\; 
\mathbf{1}\{\theta \in I(X)\}
\;=\;
\mathbf{1}\big\{L(X) \le \theta \le U(X)\big\},
\]
which is a $\{0,1\}$–valued random variable on $(\Omega,\mathcal{F},P_\theta)$. In terms of $Z$, condition \eqref{eq:coverage-def} is equivalent to
\begin{equation}
  E_\theta[Z(X)] = 1-\alpha,
\end{equation}
so that, under $P_\theta$, $Z(X)$ is Bernoulli with success probability $1-\alpha$.

Under this model, there are two natural ways of viewing the coverage variable $Z$. At the \emph{design level}, we work with its unconditional law under $P_\theta$, so that for a single use of the procedure
\begin{equation}
  P_\theta\big(Z(X) = 1\big) = 1-\alpha
  \label{eq:design-level-coverage}
\end{equation}
is a property of the pair $(P_\theta^X,I)$ and does not depend on any particular realized dataset. At the same time, $Z$ can be viewed \emph{conditional on the data} $X$. For a fixed $\theta$ and any realization $x \in \mathcal{X}$ such that $X(\omega)=x$, the interval $I(X(\omega))=I(x)$ is determined, and so is the indicator
\[
  Z(X(\omega))
  =
  \mathbf{1}\{\theta \in I(x)\}
  \in \{0,1\}.
\]
In other words, conditioning on the $\sigma$–field $\sigma(X)$ yields the degenerate conditional expectation
\begin{equation}
  E_\theta\big(Z(X)\,\big|\,X\big)
  \;=\;
  Z(X)
  \;\in\; \{0,1\}
  \quad\text{a.s.}
  \label{eq:degenerate-conditional}
\end{equation}
The two layers are linked by the usual tower property:
\begin{equation}
  1 - \alpha
  \;\stackrel{\eqref{eq:design-level-coverage}}{=}\;
  E_\theta[Z(X)]
  \;\stackrel{\text{tower}}{=}\;
  E_\theta\!\Big[ E_\theta\big(Z(X) \,\big|\, X\big)\Big],
  \label{eq:tower-coverage}
\end{equation}
where the inner expectation is the degenerate conditional from \eqref{eq:degenerate-conditional}. Thus, the design-level coverage probability $1 - \alpha$ is recovered by averaging those $\{0,1\}$ values over the sampling distribution of $X$. Both layers are internal to the same model: they are simply $Z$ viewed with respect to different $\sigma$--fields. In particular, they apply to \emph{all} intervals generated by the procedure, regardless of whether we have already observed the corresponding data or not.

\subsection{3.2 Forecasting with confidence}
\label{subsec:risk-coverage}

\subsubsection{3.2.1 Minimizing risk pre-trial}
\label{subsubsec:risk-pre}

Let $I(X) = [L(X),U(X)]$ be a $(1-\alpha)$ confidence interval for a fixed parameter value $\theta$, and recall the coverage indicator
\[
Z_\theta := \mathbf{1}\{\theta \in I(X)\}.
\]
By construction,
\[
P_\theta(\theta \in I(X)) 
\;=\; 
\mathbb{E}_\theta[Z_\theta] 
\;=\; 
1-\alpha,
\]
so $Z_\theta$ is a Bernoulli random variable with success probability $1-\alpha$ under $P_\theta$.

Before observing any data, imagine we would like to issue a probability forecast $q \in [0,1]$ for the event ``the interval will cover $\theta$,'' i.e.\ for $Z_\theta=1$. Let $S(q,z)$ be any strictly proper scoring rule for Bernoulli forecasts (for example, the Brier or log score), interpreted as a loss function in the usual sense that a lower score is better \cite{gneiting2007strictly}. From our setup, the pre-trial expected score is
\[
R_\theta(q) 
\;:=\; 
\mathbb{E}_\theta\bigl[S(q,Z_\theta)\bigr].
\]
Strict propriety of $S$ implies that $R_\theta(q)$ is uniquely minimized at
\begin{equation}
  q^* 
  \;=\; 
  P_\theta(Z_\theta=1)
  \;=\; 
  1-\alpha.
  \label{eq:qstar-pre}
\end{equation}
At the design level, then, the confidence level $1-\alpha$ is exactly the probability forecast for coverage that minimizes expected loss under any strictly proper scoring rule. In this sense, $1-\alpha$ is a model-implied \emph{predictive probability} of coverage for any single use of the procedure. By extension, any alternative constant forecast, such as ``the interval covers'' (i.e.\ $q\equiv 1$), has strictly larger expected loss whenever $0 < 1-\alpha < 1$ (Neyman might not have intended his interpretation to be used this way, but for all intents and purposes, "the interval covers" is a constant forecast about coverage, whether we as the statisticians are meant to truly "believe" the statement to be true or not).

\subsubsection{3.2.2 Minimizing risk post-trial}

After observing $X = x$, we obtain the realized interval $I(x)$ and realized indicator
\[
  Z_\theta(x) := \mathbf{1}\{\theta \in I(x)\}.
\]
Suppose now that we choose to condition on some information $\mathcal{G}$ derived from the design and the data (for example, the fact that the interval was produced by this procedure, or possibly some coarser features of $I(X)$ such as its length), but not on $\sigma(X)$ itself. A data-dependent forecast is then a random variable $q(X)$ that is $\mathcal{G}$-measurable. Its conditional expected score is
\[
  E_\theta\!\Big[S\!\big(q(X),\, Z_\theta\big)\,\Big|\,\mathcal{G}\Big].
\]
By strict propriety of $S$, this conditional risk is minimized when
\begin{equation}
  q^*(X) = E_\theta[Z_\theta \mid \mathcal{G}] = P_\theta\!\big(\theta \in I(X)\,\big|\,\mathcal{G}\big).
  \label{eq:post-trial-optimal}
\end{equation}
In the baseline case where the procedure is such that, for the chosen $\mathcal{G}$,
\begin{equation}
  E_\theta[Z_\theta \mid \mathcal{G}] = 1 - \alpha \quad \text{almost surely},
  \label{eq:baseline-conditional}
\end{equation}
the risk-minimizing forecast remains $q^*(X) = 1 - \alpha$ even after observing $I(X)$: the specific realized interval carries no further information about coverage beyond the design. In that sense, the ex-ante and (design-level) ex-post predictive probabilities coincide. By contrast, if certain features collected in $\mathcal{G}$ are known under the design to be associated with different conditional coverage, then
\[
  P_\theta\!\big(\theta \in I(X) \mid \mathcal{G}\big) \neq 1 - \alpha,
\]
and the optimal predictive probability $q^*(X)$ should be adjusted accordingly. At the other extreme, if one takes $\mathcal{G} = \sigma(X)$, the conditional expectation collapses to
\[
  E_\theta[Z_\theta \mid X] = Z_\theta \in \{0,1\},
\]
recovering the oracle's degenerate view discussed in Section~3.1. This ``forecast'' is perfect for an oracle that knows whether the interval covers, but a non-omniscient statistician who insists on the slogan ``either it covers or it does not'' is in a different position: they know only that the true coverage indicator takes values in $\{0,1\}$, but they still must name a single real number $q \in [0,1]$ when scored. In particular, the two constant forecasts consistent with the either-or stance---$q \equiv 0$ and $q \equiv 1$---both have strictly larger expected loss than $q \equiv p = 1 - \alpha$ whenever $0 < p < 1$. This follows immediately from strict propriety: $E_\theta[S(q, Z_\theta)]$ is uniquely minimized at $q = p$, and $p \notin \{0,1\}$. In this sense, the ``forecast'' implied by the either-or slogan is strictly dominated, in the usual scoring-rule sense, by the design-level forecast $1 - \alpha$.

\subsubsection{3.2.3 Design-based refinement via \texorpdfstring{$\theta$-free}{theta-free} conditional coverage}
\label{subsubsec:theta-free}

Now imagine that we have a constructed interval in hand and are wondering whether to update our pre-trial forecast from \(1-\alpha\) to something else, and, if so, to what. Since \(
\theta\) is fixed and unknown, all we have to work with is what is in the data. Let $T(X)$ be a statistic derived from the data (e.g., interval length, whether the interval hits a known boundary, or other design-specific features), and let
\[
\mathcal{G} := \sigma\bigl(T(X)\bigr)
\]
be the $\sigma$-algebra it generates. A data-dependent probability forecast is a random variable $q(X)$ that is $\mathcal{G}$-measurable. As before, we take $S(q,z)$ to be a strictly proper scoring rule for Bernoulli forecasts, interpreted as a loss. 

\begin{theorem}[Design-based optimal forecast with a $\theta$-free statistic]
\label{thm:theta-free}
Assume there exists a measurable function $g : \mathrm{range}(T) \to [0,1]$ such that, for all $\theta \in \Theta$,
\begin{equation}
  P_\theta(Z_\theta = 1 \mid \mathcal{G})
  \;=\;
  g\bigl(T(X)\bigr)
  \quad \text{almost surely.}
  \label{eq:theta-free-cond-coverage}
\end{equation}
Then the forecast rule
\[
q^*(X) := g\bigl(T(X)\bigr)
\]
uniquely minimizes the conditional expected score
\[
\mathbb{E}_\theta\bigl[ S\bigl(q(X),Z_\theta\bigr) \,\big|\, \mathcal{G} \bigr]
\]
for every $\theta \in \Theta$ and almost every realization of $T(X)$. In particular, for all $\theta \in \Theta$ and all $\mathcal{G}$-measurable $q(X)$,
\begin{equation}
  \mathbb{E}_\theta\bigl[ S\bigl(q^*(X),Z_\theta\bigr) \bigr]
  \;\le\;
  \mathbb{E}_\theta\bigl[ S\bigl(q(X),Z_\theta\bigr) \bigr],
  \label{eq:theta-free-uncond}
\end{equation}
with equality for a given $\theta$ only if $q(X) = q^*(X)$ almost surely under $P_\theta$.
\end{theorem}

Here, the condition~\eqref{eq:theta-free-cond-coverage} is a design-level statement: the conditional coverage given $T(X)$ is the same function of $T$ for all $\theta$. This is a very strong assumption, and in most real-world scenarios, it will not apply. Nonetheless, when it does, $q^*(X)=g(T(X))$ will be a uniformly risk-minimizing predictive probability for coverage across the entire parameter space, giving us a data-dependent way to update our forecast based, again, purely on the design--no prior over \(\theta\) required, because the forecast minimizes risk for them all--and in a way justified entirely by risk-minimization under repeated sampling.

\subsubsection{3.2.4 Defaulting to the confidence level}

From before, we saw that for a fixed $\theta \in \Theta$, the conditionally optimal forecast given $\mathcal{G}$ is always
\[
q_\theta^*(X) = \mathbb{E}_\theta\bigl[ Z_\theta \mid \mathcal{G} \bigr]
              = P_\theta(Z_\theta = 1 \mid \mathcal{G}).
\]
Naturally, though, if there exists no measurable $g$ such that 
\[
P_\theta(Z_\theta = 1 \mid \mathcal{G})
  = g\bigl(T(X)\bigr)
  \quad \text{for all } \theta \in \Theta,
\]
then there is no single $\mathcal{G}$-measurable rule $q(X)$ that is conditionally optimal for every $\theta$. Any non-constant $q(X)$ based on $T(X)$ will reduce the expected score for some $\theta$ and increase it for others, and choosing among such rules then necessarily requires some additional structure (e.g., a prior on $\theta$) that Neyman's machinery does not provide.

By contrast, going back to our view of the confidence level as a predictive probability, among \emph{constant} forecasts $q(X)\equiv q$, strict propriety implies that, for each fixed $\theta$,
\[
\mathbb{E}_\theta\bigl[S(q,Z_\theta)\bigr]
\]
is uniquely minimized at
\[
q = P_\theta(Z_\theta = 1) = 1-\alpha.
\]
So, again, the nominal level $1-\alpha$ is the unique constant forecast that minimizes expected loss for every $\theta$ under any strictly proper scoring rule. Moreover, because $P_\theta(Z_\theta=1)=1-\alpha$ for all $\theta$, the same conclusion holds after averaging over any prior $\pi$ on $\Theta$:
\[
\int P_\theta(Z_\theta = 1)\,\pi(d\theta) = 1-\alpha
\quad \text{for all priors } \pi.
\]
This integral holds because \(1-\alpha\) does not depend on \(\theta\), and so integrating a constant gives a constant. In this sense, when no $\theta$-free conditional coverage refinement exists, the design-level confidence $1-\alpha$ is the natural default predictive probability of coverage for a realized interval: it respects the defining calibration property for all $\theta$ and is Bayes-optimal among constant forecasts under every prior on $\Theta$ (including, again, the constant forecast implicit in always stating the interval covers). 

\subsection{3.3 Framework summary}

To recap, if we are interested in predicting whether a particular interval covers the parameter, \(1-\alpha\) is the best we can do ex ante. Ex post, if we have a \(\theta\)-free statistic that lets us break down the constant \(1-\alpha\) forecast into finer-grained conditional probabilities that still average to \(1-\alpha\), then using those finer pieces \textit{strictly} improves our prediction under any strictly proper score, unless the finer-grained conditionals are actually all the same, in which case we are effectively back at \(1-\alpha\). Whether we have such a $\theta$-free statistic will depend on the underlying model, as I outline briefly below. 

\subsubsection{3.3.1 Unbounded location–scale models}

In standard unbounded, translation-invariant models with pivot-based CIs, the coverage event is determined by a pivot, while interval width is a $\theta$-free feature of the realized interval. In such designs, neither the absolute location of the interval nor any $\theta$-free function of its geometry carries additional design-based information about coverage, and the optimal pre- and post-trial forecast for $\mathbf{1}\{\theta \in I(X)\}$ under any strictly proper scoring rule is simply the nominal level $1-\alpha$. Many familiar estimation problems will fall under this category.

\paragraph{3.3.2 Finite-window designs}

In designs where the data are supported on a finite window around $\theta$ (as in the uniform ``submarine'' model I explore in Section 4), the relative width of the interval,
\[
T(X) = \frac{\text{length}(I(X))}{\text{length of the window}},
\]
is a $\theta$-free statistic whose conditional coverage $P_\theta(\theta \in I(X)\mid T)$ typically varies with $T$. In such cases, the design-based optimal forecast is
\[
q^*(X) = P(\theta \in I(X)\mid T(X)),
\]
which strictly improves on the constant forecast $1-\alpha$ under any strictly proper score whenever the conditional coverage function is non-constant. These kinds of designs are occasionally used as instructive examples to show that adhering to the design-level coverage probability ex post can lead to apparently incoherent statements (e.g., when a $50\%$ CI ends up covering only, say, $10\%$ of the data support).

\paragraph{3.3.3 Bounded parameter spaces}

When $\theta$ is known to lie in a compact interval $[a,b]$, the normalized endpoints
\[
\tilde L = \frac{L(X)-a}{b-a}, \qquad
\tilde U = \frac{U(X)-a}{b-a}
\]
describe the interval's width and position relative to the parameter space itself. In simple symmetric designs, we can again consider the design-based conditional coverage $P_\theta(\theta \in I(X) \mid \tilde L,\tilde U)$ as a post-trial forecast. A full general theory is beyond the scope of this paper, but these normalized endpoints provide a natural starting point for simulation-based assessment of $\theta$-free refinements.

\subsubsection{3.3.4 A rough guide for forecasting}

With the foregoing in mind, we can formulate a rough step-by-step process for deciding how to make ex-post forecasts. Given a confidence procedure and a statistic $T(X)$ derived from the interval, the following rule of thumb applies:

\begin{enumerate}
  \item Check whether $T(X)$ is $\theta$-free for coverage, in the sense that $P_\theta(\theta \in I(X)\mid T)$ has the same functional form for all $\theta$;
  \item If so, and if this conditional coverage varies with $T$, use the design-based forecast $q^*(X) = P(\theta \in I(X)\mid T(X))$; and
  \item If no such $\theta$-free refinement exists, default to the nominal level $q(X)\equiv 1-\alpha$.
\end{enumerate}

In the first case mentioned above (standard unbounded, translation-invariant designs), our $\theta$-free statistics for coverage carry no more information about coverage than the design itself, and so whether we choose to condition on them or not, our ex-post forecast holds steady at $1-\alpha$. In the other two cases, they may carry extra information and, as we will see in the example below, may help us improve the quality of our predictions. In these situations, i.e., when we have reason to believe $T(X)$ is likely $\theta$-free under our design, we could then obtain usable plug-in conditional coverage probabilities by simulating datasets under a convenient $\theta$ or range of $\theta$s, tabulating coverage as a function of $T$, and then using the empirical coverage estimates as a kind of look-up table for ex post forecasts. The point here is not to solve the general problem of selecting among all possible refinements, but only to note that, when such refinements are available, they will yield forecasts that strictly improve on the best constant report of 1 - $\alpha$. 

\section{4 Returning to the lost submarine}

In their paper rebutting what they call the "Fundamental Confidence Fallacy" (FCF), Morey et al. \cite{morey2016fallacy} present a simple thought experiment to show the perils of attempting to interpret confidence intervals ex post. In short, the setup involves a surface search ship looking for a lost research sub on the ocean floor. The surface ship needs to drop a line to the sub's hatch in order to rescue its crew, so they use the pattern of bubbles coming off of the sub's hull to estimate its position relative to its own. The hatch is exactly halfway down the sub's \(10\)-meter length, and the bubbles come off of the sub uniformly at random and in pairs. The authors present a number of confidence procedures for estimating the position of the hatch, but for clarity, I will limit the analysis below to only three of them. 

\subsection{4.1 Setting things up}
\label{subsec:submarine-model}

Let $\theta \in \mathbb{R}$ denote the (unknown) horizontal location of the sub relative to the rescue vessel. Conditional on $\theta$, bubble locations are modeled as
\[
X_1, X_2 \mid \theta \;\overset{\text{i.i.d.}}{\sim}\; \mathrm{Uniform}[\theta-5,\theta+5].
\]
Write $X_{(1)} := \min\{X_1,X_2\}$, $X_{(2)} := \max\{X_1,X_2\}$, let
\[
\bar{X} := \frac{X_1 + X_2}{2},
\qquad
d := X_{(2)} - X_{(1)} = |X_1 - X_2|.
\]

We consider two $50\%$ confidence procedures for $\theta$ based on $(X_1,X_2)$ (see Morey et al.\ for derivations). The first is a nonparametric (NP) procedure that simply returns the interval spanned by the two observations:
\begin{equation}
  I^{\mathrm{NP}}(X_1,X_2)
  \;:=\;
  [X_{(1)}, X_{(2)}]
  \;=\;
  \bigl[\bar{X} - \tfrac{d}{2},\; \bar{X} + \tfrac{d}{2}\bigr].
  \label{eq:sub-np-short}
\end{equation}

The second procedure (UMP) selects, for each realization, the shorter of two intervals that have the same $50\%$ coverage under the uniform model:
\begin{equation}
  I^{\mathrm{UMP}}(X_1,X_2)
  \;:=\;
  \begin{cases}
    [X_{(1)}, X_{(2)}],
      & d < 5, \\[0.75ex]
    [X_{(2)} - 5,\, X_{(1)} + 5],
      & d \ge 5.
  \end{cases}
  \label{eq:sub-ump-short}
\end{equation}
Both $I^{\mathrm{NP}}$ and $I^{\mathrm{UMP}}$ have coverage $P_\theta(\theta \in I(X_1,X_2)) = 1/2$ for all $\theta$, but $I^{\mathrm{UMP}}$ has strictly smaller expected length. The second interval is also the likelihood of the pair of bubbles under the model.

\subsection{4.2 Simulating coverage}

To take an empirical look at coverage, I simulated $N=1e5$ runs of the experiment, at each point sampling a pair of bubbles from the distribution described above. Under this setup, we get confirmation that both procedures cover $\theta$ with the same probability: $49,998$ successes out of the $1e5$ runs yields $50\%$ coverage, rounded up. We also get confirmation that two procedures cover $\theta$ at exactly the same times, since the UMP intervals are always either equal to the NP intervals or strictly contained by them, in the latter case effectively dropping the parts of the larger interval that are inconsistent with both bubbles being within $\pm\,5m$ of the hatch. As with any CI procedure, each particular interval either does or does not contain \(\theta\), and so conditioned on their realized values, each will have a degenerate probability in $\{0,1\}$ (in the simulation, of course, we know $\theta$, so our coverage \textit{prediction} is always one of those two values and is perfectly accurate). 

\subsubsection{4.2.1 Design-level coverage as a constant forecast}

Since in the experiment the location of the hatch is unknown, we do not know how many of any particular number of constructed intervals will have included it. If we were interested solely in predicting the underlying coverage events for those intervals, what forecast would get us the best result? We can begin with the two constant forecasts mentioned earlier: the one we get from always stating that the interval covers, $q(X)=1$; and the one we get from the design-level coverage guarantee, $q(X)=1-\alpha=0.5$. Using Brier score loss as our proper scoring rule, the loss for a single forecast $q$ and outcome $z \in \{0,1\}$ is
\[
S(q,z) = (z - q)^2.
\]
From Section 3 above, we now let $Z_\theta := \mathbf{1}\{\theta \in I(X)\}$ be the coverage indicator for a single interval from either procedure. In the submarine example, $Z_\theta \sim \mathrm{Bernoulli}(p)$ with $p = 1/2$, so the expected Brier loss for a constant forecast $q$ is
\[
R_\theta(q)
:= \mathbb{E}_\theta\bigl[(Z_\theta - q)^2\bigr]
= p(1-q)^2 + (1-p)q^2.
\]
For $p = 1/2$, the two constant forecasts above yield
\[
R_\theta(1)
= \tfrac{1}{2}(1-1)^2 + \tfrac{1}{2}(1)^2
= \tfrac{1}{2},
\]
and
\[
R_\theta\bigl(\tfrac{1}{2}\bigr)
= \tfrac{1}{2}\bigl(1-\tfrac{1}{2}\bigr)^2
  + \tfrac{1}{2}\bigl(\tfrac{1}{2}\bigr)^2
= \tfrac{1}{4}.
\]
This is an obvious result, but in this setting, using the design-level forecast $q=1-\alpha=0.5$ halves the expected Brier loss relative to always asserting that the interval covers ($q=1$). Since the design-level forecast is also the optimal constant forecast, this is to be expected.

\subsubsection{4.2.2 Beating the constant forecasts with conditional coverage}

The reason Morey et al. present this thought experiment, however, is naturally not to extol the virtues of the confidence level as a constant forecast: indeed, it is primarily to show that, given the underlying model, a constant design-level forecast leads one to make rather awkward statements about coverage ex post. For example, if we use the NP procedure to construct an interval for a sample and notice that, after construction, the interval only spans $25\%$ of the $10$-meter length of the hull, should we still say it covered the hatch with $50\%$ probability? In a sense, yes, that is exactly true--as shown in Section 3, the model gives us two layers of probability that apply to each and every interval, one of them being $1-\alpha$--although it may feel unsatisfying to say, and it requires us to ignore the actual values (and relative width) of the realized interval, which throws away some information. 

Thankfully, because of the setup of experiment, with a uniform distribution of known width distributed symmetrically around $\theta$, the width of the intervals relative to the support of $X$ is in fact a $\theta$-free statistic, and so we can test it out to see whether it improves our forecast. To see these two things in action--the statistic's $\theta$-freeness and its value for prediction--I reran the simulation above across a range of values for both $\theta$ and the support (i.e., the hull width). In total, there were $100$ unique configurations, with $\theta$ (hatch location) ranging from $0$ to $10$ in increments of $1$, and scale (hull width) ranging from $10$ to $110$ in increments of $10$. As before, each experiment consisted of sampling $1e5$ pairs of bubbles from a $\mathrm{Uniform}$ distribution with the given midpoint and scale.

Table 1 shows the Brier scores for using the conditional coverage characteristics of the two estimators above to predict whether they captured the hatch, along with using the two constant forecasts of $1$ and $1-\alpha$ for the same (the scores for the constant forecasts were the same for both procedures, so they are only listed once in the table).\begin{table}
\centering
\begin{tabular}{rrr}\toprule
 Forecast& Brier score ($\mu$)& Brier score $\sigma^2$\\\midrule
 Constant $1$& $0.500$&$0.000$\\
Constant $1-\alpha$& $0.250$& $0.000$\\
NP width& $0.117$& $0.000$\\
 UMP width& $0.170$& $0.000$\\ \bottomrule
\end{tabular}
\label{Table 1}
\caption{Brier score means and variances for four different forecasting strategies when using a joint confidence procedure to estimate the location of the hatch across 100 different simulation configurations (hatch location ranging from 0 to 10 in increments of 1, and hull width ranging from 10 to 110 in increments of 10). "NP Width" denotes a forecast from coverage probability conditioned on the nonparametric interval's width relative to the support of $X$, and "UMP Width" denotes the same, but for the universally most powerful interval. Variances are shown as empirical evidence that the strategies performed consistently across the different configurations.}
\end{table}
Unsurprisingly, the conditional forecasts beat the constant design-level forecast in both cases, with the Brier score dropping from $0.25$ to about $0.12$ for the NP interval and to about $0.17$ for the UMP interval. The scores are instructive in another way as well. In this setup, the two procedures share the \emph{same} coverage indicator $Z_\theta$: by definition, whenever the NP interval covers $\theta$, the UMP interval also covers, and vice versa. The difference lies in the $\theta$-free statistic used for forecasting coverage. For the NP procedure, we condition on its relative width $D$; for the UMP procedure, we condition on its (shorter) relative width
\[
W \;=\; \min\{D,\,1-D\},
\]
which is a deterministic coarsening of $D$ that folds very wide NP intervals back into shorter UMP intervals. Thus, $W$ is a less informative function of the data than $D$ (it induces a coarser partition of the sample space), and conditional forecasts based on $D$ can only improve---and never worsen---expected loss relative to forecasts based on $W$ under any strictly proper scoring rule. In that sense, the NP interval is actually a better \emph{predictor} of coverage when we condition on relative width, even though the UMP interval is the better \emph{estimator} of $\theta$: the latter discards impossible values under the design and has shorter expected length for the same marginal coverage, but in doing so it gives up some of the $\theta$-free, coverage-relevant information carried by the wider NP interval.

Using the same conditional coverage estimates, we can also answer the question posed above about the $2.5m$-wide interval precisely: under simulation, intervals of that width cover $\theta$ approximately $33\%$ of the time and not, as Morey et al. gesture toward as a kind of paradox, $50\%$ of the time. Looking at this kind of interval, could we still forecast coverage at $1-\alpha$? Certainly, but it would be an overestimate, and we would do better by issuing the finer-grained conditional forecast instead. 

\subsubsection{4.2.3 The bit about nesting}

Morey et al. also point to nested intervals as problematic for ex-post inference about coverage, observing that if one $50\%$ CI completely contains another, then it is logically impossible for them both to cover $\theta$ each with $50\%$ probability, since there would be no probability mass left to fall in the space between them. From a frequentist standpoint, though, a natural response would be to note that once we start to reason about coverage from two CIs simultaneously, we are now talking about a \textit{composite} confidence procedure with design-level coverage defined by their joint distribution and not their individual marginals.

To see such reasoning in action, let us pick a new pair of procedures, one being the UMP procedure from above, and the other being what Morey et al. describe as the "sampling distribution" (SD) procedure, defined as $\bar{x}\pm(5-5/\sqrt{2})$. Under simulation, these have similar marginal coverage probabilities, at $\approx0.500$ for the former, and $\approx0.501$ for the latter. The SD procedure does not overlap with the UMP procedure in the same way as the NP procedure, though, and the joint coverage probability under simulation turns out to be a tad higher at $\approx0.585$. From the start, then, we know there should be no reason to consider the composite procedure as having the same chance of capturing $\theta$ as each procedure individually--the design-level coverage probability roughly $8.5\%$ higher, presumably because we are casting a wider net on average than with either procedure alone.

What, then, about nesting? Can we use it to our advantage for forecasting? As it turns out, coverage probability \textit{does} change substantially when the intervals are nested: when the SD intervals are nested inside the UMP intervals, the probability that either of them covers goes up to $0.790$, and when the roles are reversed, it drops down below the design-level joint coverage to $0.440$ (note that, because the intervals share the same midpoint by design, one is always nested inside of the other--all that changes is which one is where). These probabilities are also stable across the simulation configurations, and so they appear to be $\theta$-free. To update our ex-ante forecast from the design-level joint value $p_{\mathrm{joint}}\approx0.585$ to something different ex post, we now have two main choices: use the conditional coverage probabilities $P(Cover \,|\,\mathrm{SD}\subset \mathrm{UMP})$ and $P(Cover\,|\,\mathrm{UMP}\subset\mathrm{SD})$ given above, or use the finer-grained versions of these where use coverage probability conditioned on the relative width of whichever is the wider of the two intervals (i.e., the outer one). Table 2 shows the Brier scores for these forecasts, relative to the outcome that either of the intervals cover.
\begin{table}
\centering
\begin{tabular}{lrr}\toprule
 $\mathrm{Forecast}$&  Brier score $\mu$& Brier score $\sigma^2$\\\midrule
 Constant $1$& $0.415$&$0.000$\\
Constant $p_{\mathrm{joint}}$& $0.243$& $0.000$\\
Nest. Cond.& $0.213$& $0.000$\\
Max Width& $0.212$& $0.000$\\ \bottomrule
\end{tabular}
\caption{Brier score means and variances for four different forecasting strategies when using a joint confidence procedure to estimate the location of the hatch across 100 different simulation configurations (hatch location ranging from 0 to 10 in increments of 1, and hull width ranging from 10 to 110 in increments of 10) "Nest. Cond" denotes a forecast from coverage probability conditioned on which interval contains the other, and "Max Width" denotes a forecast from coverage probability conditioned on the relative width of whichever interval contains the other. Variances are shown as empirical evidence that the strategies performed consistently across the different configurations.}
\label{Table 2}
\end{table}
Again, a forecast of always $1$ leads to the worst loss at $0.415$; moving to the constant design-level joint forecast $q\equiv p_{\mathrm{joint}}\approx0.585$ lowers that by slightly less than half to $0.243$. Using conditional coverage probabilities under the two nesting conditions lowers a touch more to $0.213$, and combining that information with each procedure's marginal conditional coverage probability (relative to itself, not to either interval covering) leads to the lowest score at $0.212$. These results offer some mild empirical evidence in favor of the general principle raised in Section 3 that conditioning on more information leads to a better forecast. 

Before leaving the submarine example behind, we can return to the apparent logical paradox raised at the beginning of this subsection about the potentially-missing coverage probability between the inner and outer intervals in a nested pair. Across the simulations, the inner SD interval misses when the outer UMP interval covers with an average probability of $0.085$, and vice-versa (results were again stable across all combinations of $\theta$ and scale). This probability fills the "gaps" between the inner and outer nested interval in both cases and are not, thankfully, $0$, as Morey et al. had wondered whether they might be. It is also approximately the difference between either of the procedures' marginal coverage probability of $0.500$ and their joint coverage probability of $0.585$, a sensible and welcome result.

\subsubsection{4.2.4 Code availability}

Code for the lost submarine simulations is available on GitHub at \url{https://github.com/scotthlee/confidence-as-forecast}. 

\section{5 Discussion}

\subsection{5.1. The utility of confidence as a forecast}

Treating "confidence" as a probabilistic forecast gets us a few, hopefully helpful, things for free when using Neyman's machinery for inference. As an interpretation for some number of realized intervals, it gives us a principled way to estimate the number of them that covered, for example, by using $1-\alpha$ as $p$ to calculate a binomial mean. This benefit most notably applies to the case of a single realized interval, where we can now happily say both that the interval either covers, or it does not (conditioned on its endpoints), and that it has some intermediate probability of covering the parameter (conditioned on the group coverage of other intervals like it). It also allows for a straightforward interpretation of groups of intervals that very plainly would not cover the parameter with probability $1-\alpha$, since we have the mathematical support for updating our forecast based on any coverage-relevant information they may contain. As touched on above, the latter serves as a response to the numerous counter-examples in the literature intended to show that "confidence" has no coherent meaning ex post, and the former serves as formal support for Neyman's original intention for CIs as having estimable coverage properties in conducting repeated (empirical) experiments. Although these two seem the most valuable, treating "confidence" as a forecast also helps keep clear what we mean when we talk about "coverage probability", which can now be seen as both a design-level unconditional probability under the model and an information-relative conditional probability based on whatever we happen to observe when using a confidence procedure. 

\subsection{5.2 Isn't this epistemic?}

Part of why Neyman may have resisted a forecasting-based interpretation of ex-post coverage is that it threatens to blur the boundary between his frequentist, procedural, behavioristic view of inference and the then-emerging subjectivist Bayesian program of Ramsey \cite{ramsey1926truth} and de Finetti \cite{de1937prevision}. In particular, the gap between Neyman's error-controlling constructions and de Finetti's slogan that 'probability does not exist' (ibid.) may have represented a sharper break than Neyman was willing to make. Still, as the preceding sections suggest, when we treat ``confidence'' as a predictive probability for an underlying coverage event determined by the design, the resulting notion remains both objective and purely frequentist. Both ex ante and ex post, coverage probabilities are defined as limiting relative frequencies of success in clearly specified reference classes, e.g., intervals produced by this procedure, or intervals of a given relative width $W$ under a given design, and under any strictly proper scoring rule, the risk-minimizing forecast is exactly the relevant conditional coverage probability, whether unconditional ($1-\alpha$) or conditioned on a $\theta$-free statistic such as relative width. So long as those conditional probabilities are all equally well-defined under the sampling model, there seems to be no mathematical reason to treat some as more legitimate than others, or to insist that our confidence in coverage must not change, even if the design itself tells us how it should.

Taking a slightly more philosophical tack, we might also note that Neyman's behavioristic program is explicitly designed to control our long-run error rates in making statistical inferences, for example in stating that an interval covers the parameter, or in rejecting a null hypothesis. The only reason we ever make such errors, however, is that we, as agents, do not know whether the claims themselves are true. If we did---as in the familiar stacked-interval plot where $\theta$ is taken to be known, and we can see exactly which intervals cover---our error rate would be zero, and there would be no substantive need for confidence procedures in the first place. It is precisely our lack of knowledge that makes long-run error control, and hence both Neyman's framework and the error-statistical machinery built more recently around it \cite{mayo2011error}, both useful and relevant. To be sure, though, none of this is to suggest that CI theory itself is subjective in any way, or that the conditional coverage probabilities explored in Sections 3 and 4 are somehow tied to degree of belief, credence, propensity, or any other philosophical interpretation of probability aside from frequentism (again, they are defined purely with respect to limit relative frequencies under repeated sampling). Mainly, it is to suggest that the idea of "confidence", as Neyman conceived, was defined relative to an agent's (typically a statistician's) information state when making statements about unobserved events, and could thus reasonably be described as at least information-relative, if not actually epistemic in the subjective sense. 

\subsection{5.3 A note on pedagogy}

In light of this paper's analysis, it seems natural to teach students of statistics that a confidence procedure gives rise to three related ways of talking about the \emph{single-case} coverage event $\{\theta \in I(X)\}$. First, there is the degenerate $\{0,1\}$ judgment obtained by conditioning on the realized interval (or equivalently on the full data): either this particular interval covers or it does not. Second, there is the design-level probability $1-\alpha$, obtained by averaging the coverage indicator over the sampling distribution, and conditioned only on the fact that the procedure was used at all. Third, there is a predictive probability for coverage conditioned on whatever coarser information we choose to retain about the case at hand (for example, that the interval arose from a particular procedure, or that it has a certain relative width under a given design). 

Since many of the intervals students will encounter early on fall into what I called the first class of designs in Section~3.4---unbounded, approximately translation-invariant models with pivot-based CIs---it also seems fair to point out that the realized endpoints typically do not carry any \emph{design-based} information about coverage beyond the nominal level (even though they may be useful for other purposes, such as summarizing uncertainty about $\theta$ or inverting hypothesis tests). In such settings, the best forecast for coverage, both before and after seeing the interval, is simply $1-\alpha$.

With this perspective in mind, it may be pedagogically helpful to introduce the \emph{general} notion of a confidence procedure first, and only then show how it specializes to the canonical introductory intervals, such as the $t$ interval for a normal mean, the Wald interval for a binomial proportion, or nonparametric bootstrap intervals. As Basu and others have emphasized \cite{dasgupta2010ancillary}, it is even possible to construct confidence procedures based entirely on randomization variables whose distributions are themselves $\theta$-free, without specifying any sampling model for $X$ given $\theta$. His example, together with the uniform-design procedures of Welch \cite{welch1939confidence} and its reimagining in the submarine example of Morey et al.\ \cite{morey2016fallacy}, form a useful constellation of procedures for helping students see that ``confidence'' is fundamentally about the coverage indicator as a Bernoulli-distributed event under a specified design, rather than primarily about pinning down an approximate value of $\theta$ (even though the latter is often the motivating goal in applications).

\section{6 Conclusion}

In this paper, I framed the frequentist notion of "confidence" as a probabilistic forecast that can be estimated, updated, and scored relative to the corresponding intervals' underlying coverage events. A simple thought experiment showed how we might wish to use the design-level unconditional, rather than the degenerate conditional, coverage probability for betting, and the following section develop that notion formally by treating coverage as Bernoulli and "confidence" as a prediction whose quality can be measured with a strictly proper scoring rule. I then showed how this framework helps resolve one well-cited example of the apparent paradoxes inherent to CI-based inference, and ended by discussing the framework's frequentist bona fides, as well as its potential to reframe how we teach CI theory to beginning statistics students.

\printbibliography
\newpage
\section{Supplemental Methods}

\subsection{1 Proof sketch for theta-free conditional coverage}

Fix $\theta \in \Theta$. By strict propriety of $S$ in the conditional sense, for any $\mathcal{G}$-measurable $q(X)$,
\[
\mathbb{E}_\theta\bigl[ S\bigl(q(X),Z_\theta\bigr) \,\big|\, \mathcal{G} \bigr]
\]
is minimized (almost surely) when
\[
q(X) = \mathbb{E}_\theta\bigl[ Z_\theta \mid \mathcal{G} \bigr]
     = \mathbb{P}_\theta(Z_\theta = 1 \mid \mathcal{G}).
\]
Under assumption (21) in the main text, this conditional expectation equals $g\bigl(T(X)\bigr)$ for every $\theta$, so $q^*(X) = g(T(X))$ solves the conditional optimization problem simultaneously for all $\theta$. Taking expectations over $\mathcal{G}$ yields the unconditional risk inequality, with uniqueness from strict propriety.
\hfill$\square$

\end{document}